\documentclass[%
 reprint,
superscriptaddress,
showpacs,
 amsmath,amssymb,
 aps
]{revtex4-1}

\usepackage{graphicx}
\usepackage{dcolumn}
\usepackage{bm}
\usepackage{color}

\begin{document}

\preprint{hogehoge}

\title{Implementation of the multiconfiguration time-dependent Hatree-Fock
method for general molecules on a multi-resolution Cartesian grid}

\author {Ryohto Sawada}
\affiliation{Department of Applied Physics, Graduate School of Engineering, The University of Tokyo, 7-3-1 Hongo, Bunkyo-ku, Tokyo 113-8656, Japan}
\affiliation{Photon Science Center, Graduate School of Engineering, The University of Tokyo, 7-3-1 Hongo, Bunkyo-ku, Tokyo 113-8656, Japan}
\author{Takeshi Sato}
\affiliation{Photon Science Center, Graduate School of Engineering, The University of Tokyo, 7-3-1 Hongo, Bunkyo-ku, Tokyo 113-8656, Japan}
\affiliation{Department of Nuclear Engineering and Management, Graduate School of Engineering, The University of Tokyo, 7-3-1 Hongo, Bunkyo-ku, Tokyo 113-8656, Japan}
\author{Kenichi L. Ishikawa}
\affiliation{Photon Science Center, Graduate School of Engineering, The University of Tokyo, 7-3-1 Hongo, Bunkyo-ku, Tokyo 113-8656, Japan}
\affiliation{Department of Nuclear Engineering and Management, Graduate School of Engineering, The University of Tokyo, 7-3-1 Hongo, Bunkyo-ku, Tokyo 113-8656, Japan}
\date{\today}

\begin{abstract}

We report a three-dimensional numerical implementation of multiconfiguration time-dependent Hartree-Fock (MCTDHF)
 based on a multi-resolution Cartesian grid, with no need to assume any
 symmetry of molecular structure. We successfully compute high-harmonic generation (HHG) of ${\rm H}_2$ and ${\rm H}_2{\rm O}$. The present implementation will open a way to the first-principle theoretical study of intense-field and attosecond-pulse induced ultrafast phenomena in general molecules.

\end{abstract}

\maketitle


\section{Introduction}
The dynamics of atoms and molecules under intense (typically $\gtrsim
10^{14}\,{\rm W/cm}^2$) laser pulses is of great interest in a variety
of fields such as attosecond science and high-field physics
\cite{Krausz2009RMP,Levis2001,Calegari2014}, with a goal to directly
measure and manipulate electronic motion. Numerical simulations of such electron dynamics are a challenging task \cite{Ishikawa2015JSTQE}. Direct solution of the time-dependent Schr\"odinger equation (TDSE) cannot be applied beyond He, ${\rm H}_2$, and Li, due to a prohibitive computational cost. Thus, one of major recent directions attracting increasing interest is the multiconfiguration self-consistent-field (MCSCF) approach, which expresses the total wave function $\Psi (t)$ as a superposition \cite{Nguyen-Dang2009CP,Nguyen-Dang2013JCP,Miranda2011JCPa,Sato2013,Sato2015,Alon2007}
\begin{equation}
\label{eq:wfbe}
|\Psi (t) \rangle = \sum_{J}c_{J}(t) |J \rangle ,
\end{equation}
of Slater determinants $|I \rangle$ built from the spin orbitals $|\phi_{(i,\sigma)}\rangle = |\phi_{i} \rangle \otimes |\sigma \rangle$, where $\{ \phi_i \}$ and $\sigma \in \{\alpha,\beta \}$ denote one-electron spatial orbital functions and spin eigenfunctions, respectively. Different variants with this {\it ansatz} have recently been actively developed \cite{Ishikawa2015JSTQE}.

The time-dependent configuration-interaction (TDCI) methods take the orbital functions to be time-independent and propagate only CI coefficients $c_I(t)$. Santra {\it et al.} \cite{Greenman2010} have implemented its simplest variant, i.e., the time-dependent configuration-interaction singles (TDCIS) method to treat atomic high-field processes. In this method, only up to single-orbital excitation from the Hartree-Fock (HF) ground-state is included. Bauch {\it el al.} \cite{Bauch2014} have recently developed TD generalized-active-space CI based on a general CI truncation scheme and discussed its numerical implementation for atoms and diatomic molecules.

In the other class of MCSCF approaches, not only CI coefficients but also orbital functions are varied in time. The multiconfiguration time-dependent Hartree-Fock (MCTDHF) \cite{Caillat2005,Kato2004} considers all the possible electronic configuration for a given number of spin orbitals. As its flexible generalizations, we have recently formulated the TD complete-active-space self-consistent field (TD-CASSCF) \cite{Sato2013} and TD occupation-restricted multiple-active space (TD-ORMAS) \cite{Sato2015} methods. The latter is valid for general MCSCF wave functions with arbitrary CI spaces \cite{Ishikawa2015JSTQE,Haxton2015PRA} including, e.g., the TD restricted-active-space self-consistent-field (TD-RASSCF) theory developed by Miyagi and Madsen \cite{Miyagi2013PRA}. Numerical implementations of MCTDHF for atoms as well as diatomic molecules have been reported for the calculation of valence and core photoionization cross sections \cite{Haxton2012PRA}. We have also implemented TD-CASSCF for atoms by expanding orbital functions with spherical harmonics and successfully computed high-harmonic generation and nonsequential double ionization of Be {\color{black} \cite{Sato2015-3D}}.



Practically all the existing implementations are intended for atoms and diatomic molecules, exploiting the underlying symmetries with either  {{\color{black} the spherical \cite{Parker1996,Smyth1998,Ishikawa2005,Ishikawa2012PRL,Ishikawa2013}, cylindrical \cite{Harumiya2000,Harumiya2002,Ohmura2014,Ohmura2014b}, or prolate spheroidal \cite{Guan2010,Tao2009,Tao2009a,Tao2010}} coordinates.

In this study, we report a three-dimensional (3D) numerical
implementation of MCTDHF based on a multi-resolution Cartesian grid, with no need to assume any symmetry of molecular structure, this can in principle be applied to any molecule. With the use of a multi-resolution finite-element representation of orbital functions, we can fulfill a high degree of refinement near nuclei and, at the same time, a simulation domain large enough to sustain departing electrons. As demonstrations, we successfully compute high-harmonic generation (HHG) from ${\rm H}_2$ and ${\rm H}_2{\rm O}$. The present implementation will open a way to the first-principle theoretical study of intense-field and attosecond-pulse induced ultrafast phenomena in general molecules.

This paper is organized as follows. In Sec.\ \ref{sec:MCTDHF}, we
briefly summarize the MCTDHF method. Section \ref{sec:multi-resolution}
describes the multi-resolution cartesian grid. Section
\ref{sec:implementation} explains the numerical procedure that we implement. In
Sec.\ \ref{sec:sim}, we show examples of simulation results for He, ${\rm H}_2$, and ${\rm H}_2{\rm O}$. Conclusions are given in Sec.\ \ref{sec:conc}. Atomic units are used throughout unless otherwise stated.


\section{MCTDHF}
\label{sec:MCTDHF}


In the MCTDHF method \cite{Caillat2005,Kato2004}, the sum in Eq.\,(\ref{eq:wfbe}) runs over the
complete set of $\binom{M}{N_{\alpha}}\binom{M}{N_{\beta}}$ Slater determinants $|J \rangle$
that can be constructed from $N_{\alpha}$ electrons with spin-projection
$\alpha$, $N_{\beta}$ electrons with spin-projection $\beta$, and $M$
spatial orbitals. Their spin-projection is consequently restrited to
$S_{z}=(N_{\alpha}-N_{\beta})/2$.

Let us consider a Hamiltonian in the length gauge,
\begin{eqnarray}
\label{eq:ham}
H(t)=H_{1}(t)+H_{2},
\end{eqnarray} 
\begin{eqnarray}
\label{eq:h1}
H_{1}(t)=\sum_{i=1}^{N} \left(-\frac{\nabla_i^{2}}{2} -\sum_{a}
\frac{Z_{a}}{\left| {\bf x}_{i}-{\bf X}_{a} \right|}+{\bf x}_{i}\cdot
{\bf E}(t)\right), 
\end{eqnarray}
 \begin{eqnarray}
\label{eq:h2}
H_{2}=\sum_{i=1}^{N} \sum_{j=1}^{i-1} \frac{1}{|{\bf x}_{i}-{\bf x}_{j}|},
\end{eqnarray}
where $N=N_\alpha + N_\beta$, $X_{a}$ and $Z_{a}$ are the charge and position of the $a$-th
atom, respectively, and {\bf E}(t) is the laser electronic field. One can derive
the equations of motion for the CI coefficients $c_J(t)$ and spatial orbital functions $\phi_i (t)$, resorting to the time dependent variational principle \cite{Dirac1930,Frenkel1934,Kramer1981},
\begin{eqnarray}
\label{eq:df}
\delta \left( \int_{t_{1}}^{t_{2}} \langle \Psi | H(t)-i\partial_{t} | \Psi
	\rangle  dt \right)=0
\end{eqnarray} 
with additional constraints for uniqueness \cite{Caillat2005},

\begin{eqnarray}
\label{eq:uni}
\left\langle \phi_{j} | \phi_{k}  \right\rangle=\delta_{j,k},\ \ \ \left\langle \phi_{j} | \frac{\partial \phi_{k}}{\partial t} \right\rangle =0.
\end{eqnarray} 
The equations of motion are,
\begin{eqnarray}
\label{eq:aj}
i\dot{c}_{J} = \sum_{K} \left\langle J |H(t)| K \right\rangle c_{K},
\end{eqnarray} 
and
\begin{eqnarray}
\label{eq:phi}
i|\dot{\phi}_{i}\rangle &=& \hat{P} \left( H_{1}(t)|\phi_{i}\rangle +
					       \sum_{jklm}
					       (\rho^{-1})_{ij}
					       \rho^{(2)}_{jklm}\hat{g}_{lm}
					       |\phi_{k}\rangle \right
					       ), \nonumber \\
\end{eqnarray} 
with,
\begin{align}
\label{eq:defsp}
\hat{P}&= \hat{\bf 1}-\sum_{j=1}^{M} |\phi_{j}\rangle \langle \phi_{j}|, \\
\label{eq:defsrho}
\rho_{i,j} &=  \sum_{\sigma} \langle \Psi | \hat{a}_{i\sigma}^{\dagger}  \hat{a}_{j\sigma} | \Psi \rangle, \\
\label{eq:defsrho2}
\rho^{(2)}_{jklm} &=  \sum_{\sigma \tau} \langle \Psi | \hat{a}_{j\sigma}^{\dagger} \hat{a}_{l\tau}^{\dagger}  \hat{a}_{m\tau} \hat{a}_{k\sigma} | \Psi \rangle, \\
\label{eq:defsg}
g_{lm}({\bf x}) &= \int d{\bf x}^{'} \phi^{*}_{l}({\bf x}^{'})
\frac{1}{|{\bf x}-{\bf x}^{'}|} \phi_{m}({\bf x}^{'}),
\end{align}
where $\hat{\bf 1}$ denotes the identity operator, and $\hat{a}_{i\sigma}^{\dagger}$ and $\hat{a}_{i\sigma}$ the Fermion creation and annihilation operators, respectively, associated with spatial orbital $i$ and spin $\sigma$.
Equation (\ref{eq:defsg}) is computed by solving the Poisson equation,
\begin{eqnarray}
\label{eq:poisson}
\nabla^{2} g_{lm}({\bf x}) = -4\pi \phi^{*}_{l}({\bf x}) \phi_{m}({\bf x}).
\end{eqnarray} 
It is convenient to rewrite Eq.~(\ref{eq:phi}) as,
\begin{eqnarray}
\label{eq:wfphi}
i\dot{\phi}_{i} = \hat{P}\left( T\phi_{i} +W_{i}(t) \right)
\end{eqnarray}
where $T$ is kinetic energy.

\section{multi-resolution cartesian grid}
\label{sec:multi-resolution}

We discretize spatial orbital functions on a multi-resolution Cartesian grid,
inspired by the work of Bischoff and Valeev \cite{Bischoff2011} and the finite volume method \cite{Versteeg2007}.
Figure \ref{fig:grids} (a) schematically shows how to generate it. We start from an equidistant Cartesian grid composed of cubic cells. 
If a given cell is too large to represent orbital functions with sufficient accuracy, typically near the nuclei, 
we subdivide it into eight cubic cells with half the side length of the original cell. We continue the subdivision 
until accuracy requirements are satisfied. The center of each cube is taken as the grid point representing the cell.

The Laplacian $\nabla^2\phi$ of orbital function $\phi ({\bf r},t)$ is evaluated at each grid point by finite difference. We first illustrate it for a one-dimensional case for simplicity in Fig.~\ref{fig:grids}(b). One can evaluate the second derivative of a function $f(x)$ at grid point $x_i$ as,
\begin{equation}
\frac{d^2}{dx^{2}}f(x_{i}) \approx \frac{g_{i}^{+}-g_{i}^{-}}{\Delta x_{i}},
\end{equation}
where $\Delta x_i$ is the size of cell $i$, and the first derivatives $g_{i}^{\pm}$ at the cell boundaries are approximated by, 
\begin{align}
\label{eq:1ddiff}
g_{i}^{+} &\approx \frac{2[f(x_{i+1})-f(x_{i})]}{\Delta x_{i}+\Delta x_{i+1}},\\
g_{i}^{-} &\approx  \frac{2[f(x_{i})-f(x_{i-1})]}{\Delta x_{i}+\Delta x_{i-1}}.
\end{align}

We show the extension to two dimensions in Fig.~\ref{fig:grids}(c). The grid
points are marked by red and blue circles. In order to evaluate the
second derivative with respect to the vertical direction at the center
of cell $i$, we need the first derivative evaluated at the cell boundary
marked by the orange triangle, for which we need, in turn, the value of the function at the position marked by the star in cell $i+1$. We approximate this latter by the value $f_{i+1} \equiv f({\bf r}_{i+1})$ at the grid point, i.e., the center of the cell $i+1$. Though inferior in terms of accuracy, this scheme is much more advantageous in terms of computational cost over conventional methods such as the alternating direction implicit method \cite{Peaceman1955}, moving least squares \cite{Levin1998,Lopreore1999}, and symmetric smoothed particle hydrodynamics \cite{Batra2007,Tsai2012}.  

Then, in the 3D case, we evaluate the Laplacian $\nabla^2\phi ({\bf r})$ as,
\begin{equation}
	\nabla^2\phi ({\bf r}_a) \approx L_{aa}\phi ({\bf r}_a) + \sum_{b}{}^\prime L_{ab} \phi ({\bf r}_b),
\end{equation}
where $a$ and $b$ are cell indices, the primed sum is taken over the cells adjacent to the $a$-th cell, and,
\begin{align}
\label{eq:mctdhf-mat2ag-1}
L_{aa} &= - \sum_{b}{}^\prime L_{ab}, \\
\label{eq:mctdhf-mat2ag-2}
L_{ab} & =\frac{l_{b}^{2}}{l_{a}^{2}}\frac{2}{l_{a}+l_{b}}\frac{1}{l_{a}} \qquad (a \ne b \mbox{ and } l_{b}<l_{a}), \\
\label{eq:mctdhf-mat2ag-3}
L_{ab} & =\frac{2}{l_{a}+l_{b}}\frac{1}{l_{a}} \qquad (a \ne b \mbox{ and } l_{b} \geq l_{a}),
\end{align}
with $l_{a}$ being the side length of the $a$-th cell. 
{\color{black}
If $l_b < l_a$, a face of a cell of side length $l_a$ would contact with
$l_a^2 /l_b^2$ adjacent cells of side length $l_b$. The prefactor
$l_b^2/l_a^2$ of Eq. (\ref{eq:mctdhf-mat2ag-2}) takes into account the weight of each of the
latter.}



\begin{figure*}[t]
\includegraphics[width=16cm, bb = 0 0 607 390 ]{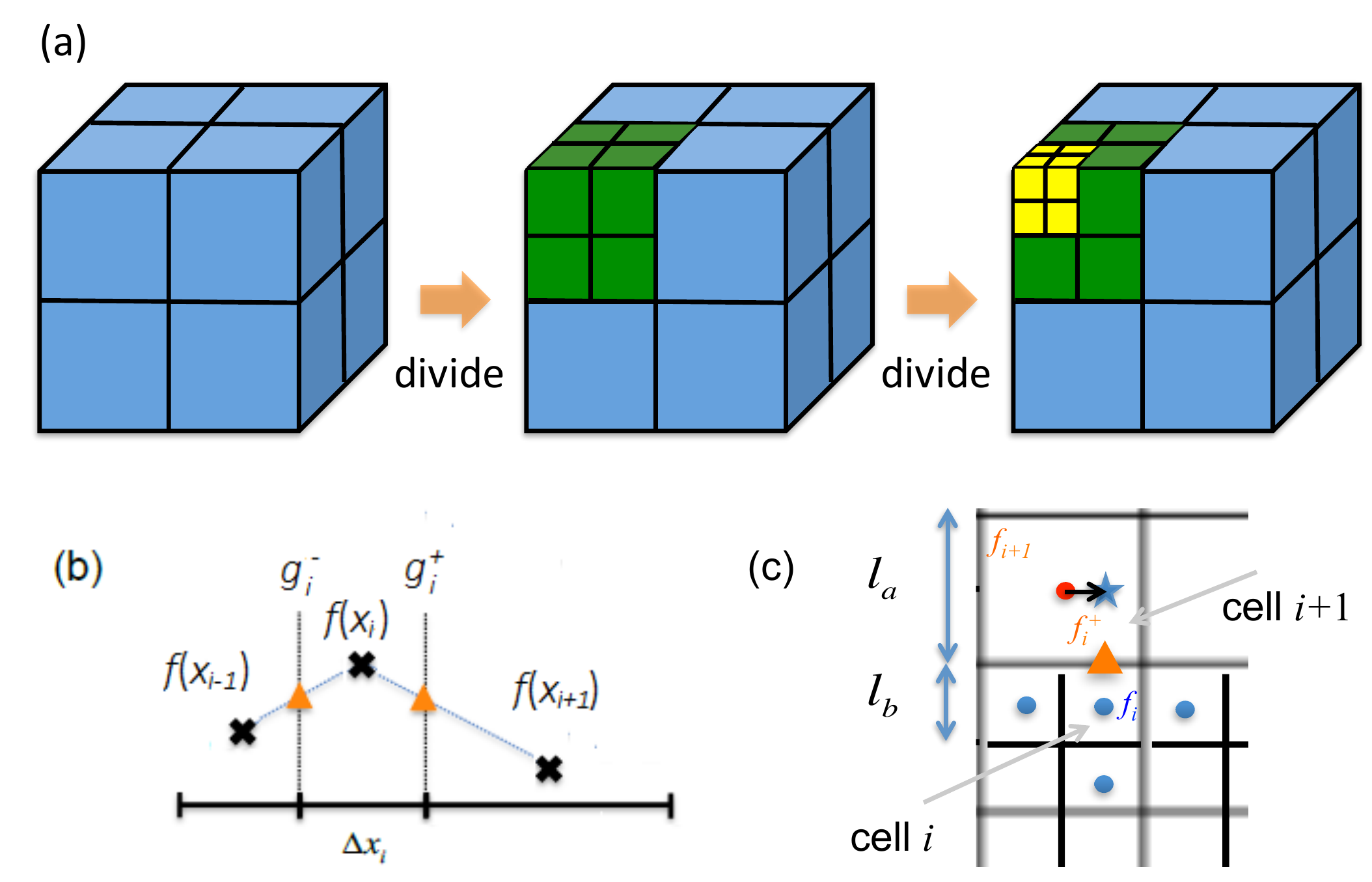}
\caption{(a) Schematic of the cartesian-based multi-resolution grids. (b)
 Schematic of the computation of second order differential at one-dimensional
 irregular grids. (c) Schematic of the computation of the first differential at the
 surface of the grid at two-dimensional irregular grids. Real grid points
 are red and blue circles.}
\label{fig:grids}
\end{figure*}

\section{Numerical procedure}
\label{sec:implementation}
We present the essential steps of MCTDHF simulations using multi-resolution cartesian grid as follows:\\

\noindent
{\it Step 1: Generation of grid and Laplacian matrix} \\

We consider a cuboid simulation region $V$ centered at the origin:
$\{{\bf x}=(x,y,z)\in \mathbb{R}^2\quad |\quad x\in [-x_L,x_L], y\in
[-y_L,y_L], z\in [-z_L,z_L], x_L>0, y_L>0, z_L>0 \}$. We set the
locations of the grid points and prepare the Laplacian matrix elements $L_{ab}$ using Eqs.~(\ref{eq:mctdhf-mat2ag-1})--(\ref{eq:mctdhf-mat2ag-3}). These are done only once in the beginning.\\

\noindent
{\it Step 2: Computation of $\rho$ and $\rho^{(2)}$}\\

Each time step starts with the computation of $\rho$ and $\rho^{(2)}$, using Eqs.~(\ref{eq:defsrho}) and (\ref{eq:defsrho2}), respectively.\\

\noindent
{\it Step 3: Computation of $g_{lm}$} \\

We solve the Poisson equation (\ref{eq:poisson}) to obtain $g_{lm}$, by
the conjugate residual method
\cite{Steiefel1955,Yousef2003}. {\color{black} The condition at the
simulation boundary $\partial V$ is given by the multipole expansion 
\begin{align}
	g_{lm}({\bf x}_{\rm bound})&=\int_V \frac{1}{|{\bf x}_{\rm bound}-{\bf x}'|} \phi_{l}^{*}({\bf x}') \phi_{m}^{*}({\bf x}')  d{\bf x}' \\
\label{eq:bound}
&=\sum_{l=0}^{\infty} \int_V \frac{|{\bf x}'|^{l} P_{l}(\cos \theta)}{|{\bf x}_{\rm bound}|^{l+1}}\phi_{l}^{*}({\bf x}') \phi_{m}^{*}({\bf x}')  d{\bf x}'
\end{align}
for ${\bf x}_{\rm bound} \in \partial V$ where {\color{black}{} $P_{l}(z)$ denotes the Legendre polynomial, and $\theta$ the angle between ${\bf x}'$ and ${\bf x}_{\rm bound}$. In the present study, we truncate the sum in Eq.~(\ref{eq:bound}) at $l=2$ (second-order multipole expansion).}} \\

\noindent
{\it Step 4: Time propagation of $c_{J}$ and $\phi_{i}$} \\

We solve the equations of motion Eqs.~(\ref{eq:aj}) and (\ref{eq:wfphi}) using a second-order exponential integrator \cite{Cox2002,Bandrauk2013}. Equation (\ref{eq:aj}) is integrated as,
\begin{align}
\label{eq:cjho}
c_{J}^{(1)}(t+\Delta t) &=c_{J}(t) + \Delta t \sum_{K} \langle J |H |K \rangle c_{K}, \\
c_{J}^{(2)}(t+\Delta t) &= c^{(1)}_{J}(t+\Delta t) \nonumber\\
&+ \Delta t \sum_{K} \langle J^{(1)} |H |K^{(1)} \rangle c^{(1)}_{K}(t+\Delta t), \\
c_{J}(t+\Delta t) &= \frac{c_{J}(t)+c_{J}^{(2)}(t+\Delta t)}{2},
\end{align}
where $|J^{(1)}\rangle$ with superscript ``(1)'' denotes the Slater determinant constructed with orbital functions $\phi^{(1)}_{i}$ defined below in Eq.~(\ref{eq:phi_i-(1)}). Equation (\ref{eq:wfphi}) is integrated as,

{\color{black}{}
\begin{align}
\label{eq:phi_i-(1)}
\phi_{i}^{(1)}& =\phi_{i}(t)\nonumber \\
&+ \hat{P}\frac{1}{1+i\Delta t T/2} \left[ (-i\Delta t T)\phi_{i}(t) +\Delta t W_{i}(t) \right],
\end{align}
\begin{align}
&\phi_{i}^{(2)}(t+\Delta t)=\phi_{i}^{(1)} \nonumber \\
&+ \hat{P}^{(1)}\frac{1}{1+i\Delta t T/2} \left[ (-i\Delta t
						 T)\phi_{i}^{(1)} +\Delta t
						W_{i}(t+\Delta t) \right], 
\label{eq:expoho}
\end{align}
\begin{align}
\hat{P}^{(1)}&=\hat{\bf 1}-\sum_{j=1}^{M} |\phi_{j}^{(1)}\rangle \langle \phi_{j}^{(1)}|, 
\end{align}
\begin{align}
\phi_{i}(t+\Delta t) &= \frac{\phi_{i}^{(2)}(t+\Delta t)+\phi_{i}(t)}{2}.
\end{align}
In Eqs.\ (\ref{eq:phi_i-(1)}) and (\ref{eq:expoho}), $(1+i\Delta t T/2)^{-1}$ is operated by the conjugate residual method \cite{Steiefel1955,Yousef2003}}. \\

\noindent
{\it Step 5a: Absorbing boundary (only in real time propagation)} \\

To prevent the reflection from the grid boundaries, after each time step, $\phi_{i}$
is multipled by a cos mask function $M(x,y,z)$ that varies from 1 to 0 between
the absorption boundary set at $x=x_0$, $y=y_0$, and $z=z_0$ ($0<x_0<x_L, 0<y_0<y_L, 0<z_0<z_L$) and the outer boundary $\partial V$ 
\cite{Krause1992,Beck2000}: 
\begin{eqnarray}
\label{eq:mask}
M(x,y,z) = C\left( \frac{|x|-x_{0}}{x_{L}-x_{0}} \right) C\left(
	     \frac{|y|-y_{0}}{y_{L}-y_{0}} \right) C\left(
	     \frac{|z|-z_{0}}{z_{L}-z_{0}} \right) \nonumber \\
\end{eqnarray}
where 
\begin{eqnarray}
\label{eq:mask2}
C(x) = 1\ \ (x \le 0),\ \  \cos(x)\ \ (x>0) .
\end{eqnarray}
Alternatively, one may use, e.g., exterior complex scaling  \cite{Scrinzi1998,Scrinzi2010}.\\

\noindent
{\it Step 5b: Rescaling of $c_{J}$ and orthonormalization of 
  $\phi_{i}$ (only in imaginary time propagation)} \\

We obtain the initial ground state via the imaginary time propagation \cite{Kosloff1986}. After each (imaginary) time step, $c_{J}$ is rescaled so that $\sum_{J}|c_{J}|^{2}=1$, and $\phi_{i}$ is orthonormalized through the Gram-Schmidt algorithm. \\

\noindent
{\it Step 6: End of time step} \\

We go back to Step 2 to start next time step.

\section{Examples}
\label{sec:sim}
\subsection{Benchmark : HHG from helium}
\label{sec:heb}

We simulate the HHG from a helium atom located at the origin. 
The side length of the cell is set to be 0.6 ($r>4$), 0.3 ($2<r<4$) and 
0.15($r<2$) respectively, depending on the distance $r$ of the grid point at the center of each cell and the origin. We also set $x_{L}=70,y_{L}=z_{L}=35$ and $x_{0}=0.7x_L,y_{0}=0.7y_L,z_{0}=0.7z_L$. The time step size $\Delta t$ is set to be 0.0025.
We consider a laser pulse linearly polarized along the $x$ axis, whose electric field $E(t)$ is given by,
\begin{align}
\label{eq:lasp2}
E(t)&=E_{\rm env}(t)\sin \omega t, \\ 
E_{\rm env}(t)&=
\left\{
\begin{array}{ll}
	\omega t/2\pi & (\omega t<2\pi), \\
	2-\omega t / 2\pi & (2\pi <\omega t<4\pi), \\
	0 & (\mbox{otherwise}),
\end{array}
\right.
\end{align}
with a central wavelength of 400 nm and a peak intensity of $8 \times 10^{14}$ W/cm$^{2}$.
For such an ultrashort pulse, the cutoff energy predicted by the semiclassical three step model {\color{black} \cite{Corkum1993,Kulander1993}} is {\color{black} $I_{p}+2.07U_{p} = 49.6\, \mbox{eV}$}, which corresponds to the {\color{black} 16.0 th} order {\color{black} where $I_p$ is ionization potential and $U_p$ is pondermotive energy.} The harmonic spectrum is obtained from the Fourier transform of the dipole acceleration. 

In Fig.\ \ref{fig:hhghe} we compare the HHG spectrum calculated with the present implementation with that calculated with another implementation in spherical coordinates {\color{black} \cite{Sato2015-3D}}. One can see that they agree with each other very well.

\begin{figure}[tb]
\includegraphics[width=8cm, bb = 0 0 352 201]{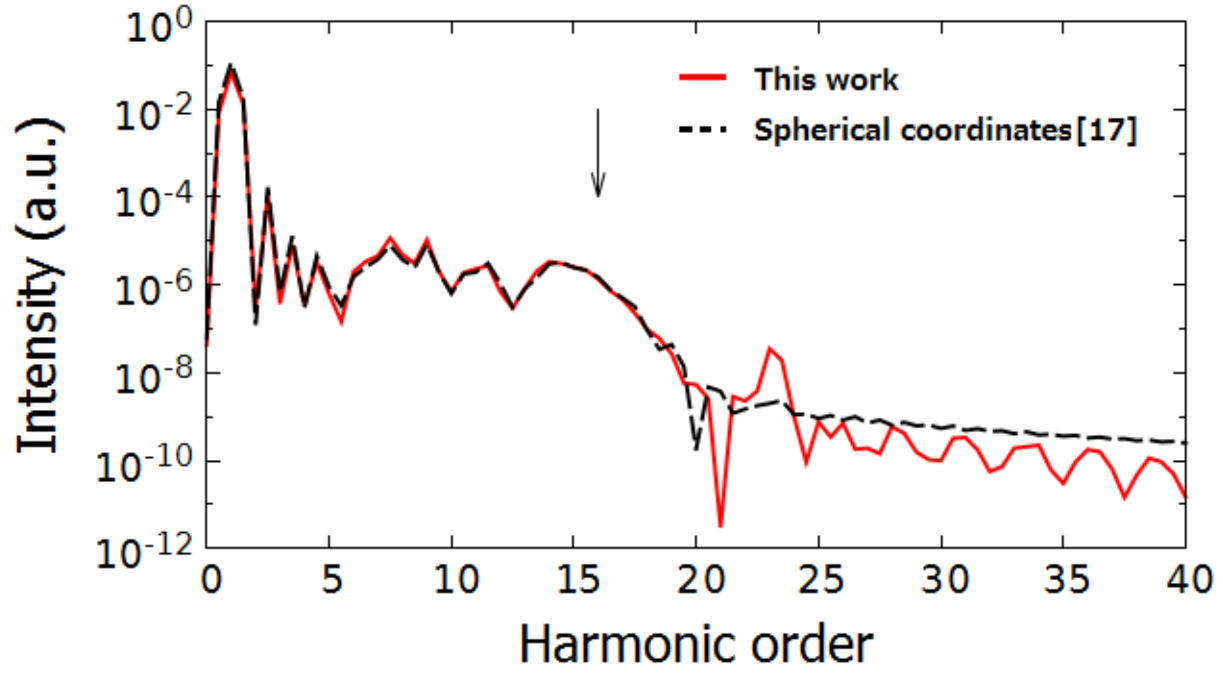}
\caption{HHG spectrum of helium computed by multi-resolution
 MCTDHF(red-solid) and method of \cite{Sato2015-3D}(black-dashed). The arrow represents the cutoff energy.}
\label{fig:hhghe}
\end{figure}

\subsection{HHG from a hydrogen molecule}
Next, we simulate the HHG from molecular hydrogen {\color{black} where {\color{black}{} two hydrogen atoms are located at $(\pm 0.7,0,0)$, respectively.} }
The side length of cell is set to be {\color{black} $l_{0}$ ($r_{0}>4$), $l_{0}/2$ ($2<r_{0}<4$)
and $l_{0}/4$($r_{0}<2$), respectively where $l_{0}$ is {\color{black}{} the side length of the largest cells (see Table \ref{table:ene} for its values).}} 
We also set $x_{L}=y_{L}=z_{L}=27$ and $x_{0}=0.7x_L,y_{0}=0.7y_L,z_{0}=0.7z_L$. The time step size $\Delta t$ is set to be 0.01. 

The ground-state energy, obtained through relaxation in imaginary time, is shown in Table\ \ref{table:ene} where {\color{black}$M$}
is the number of orbitals. It consistently tends to the literature value -1.8884 a.u. {\color{black} \cite{Turbiner2007}} {\color{black}{} with an increasing number of orbitals}. The slight {\color{black}{} dependence on $M$ and $l_0$} has only a small impact on calculated harmonic spectra, as we will see below in Fig.\ \ref{fig:hhgh2}(a,b). {\color{black}{} The values in column labeled ``0.7$^{*}$" are obtained with grids displaced parallel to the $x$ axis by 0.025. One can see that the resulting loss of grid symmetry with respect to the $yz$ plane also has only a small impact.}

\begin{table}[tb]
  \begin{tabular}{|c|r|r|r|r|}
\hline
number of & \multicolumn{4}{c|}{{\color{black} largest cell side length $l_{0}$}}\\ \cline{2-5}
orbital $M$ & 0.7 & 0.7$^{*}$ & 0.6 & 0.55 \\
\hline
1 & -1.83661 &-1.83622 & -1.84318& -1.84123\\
\hline
2 & -1.85451 &-1.85467 & -1.86164 & -1.85964\\
\hline
3 & -1.86218 & -1.86233 & -1.86925 & -1.86723 \\
\hline
6 & -1.87329 & -1.87342 & -1.88027 & -1.87756\\
\hline
  \end{tabular}
\caption{\label{table:ene} {\color{black}{} Ground-state energy (a.u.) of a hydrogen molecule, obtained by relaxation in imaginary time. The values in column labeled ``0.7$^{*}$" are obtained with grids displaced parallel to the $x$ axis by 0.025 a.u.}}
\end{table}

Let us consider a linearly polarized laser pulse with a central wavelength of 800 nm, a peak intensity of $1 \times 10^{14}$ W/cm$^{2}$, and an eight-cycle sine-squared envelope,
\begin{equation}
\label{eq:lasp1}
E(t)=E_{0}\sin^{2}(\omega t/16) \cos (\omega t).
\end{equation}
{\color{black}{} Figure \ref{fig:hhgh2} presents the HHG spectra for laser
polarization parallel to the molecular axis (the $x$ axis) [Fig.\
\ref{fig:hhgh2}(a)(b)] and 30 degrees from the molecular axis [Fig.\
\ref{fig:hhgh2}(c)]. The cutoff energy predicted by the semiclassical
three step model is 34.3\, \mbox{eV}, which corresponds to order 22.1.} One can see that the simulation is converged with respect to {\color{black}{} the number of orbitals [Fig.\ \ref{fig:hhgh2}(a)] and grid spacing [Fig.\ \ref{fig:hhgh2}(b)]}. Our multi-resolution Cartesian-grid MCTDHF, with no a priori assumption of symmetry, can also handle laser polarization oblique to the molecular axis {\color{black}{} [Fig.\ \ref{fig:hhgh2}(c)]}. 

In Fig.\ \ref{fig:hhgh2} we can clearly see the second plateau, somewhat weaker than the first one, extending beyond the cutoff ($\sim$ order 22.1). The second cutoff position is consistent with the value (53.6\, \mbox{eV} or the 34.6-th order) predicted by the three step model with the ionization potential of H$_{2}^{+}$ (34.7\, \mbox{eV}). Hence, based on a speculation that the second plateau harmonics are generated from H$_{2}^{+}$ produced via strong-field ionization, we have simulated the HHG from this molecular ion with the same laser parameters. The obtained harmonic spectrum multiplied with the ionization probability of H$_{2}$ ($2.4\times 10^{-4}$) is plotted as a yellow dashed line in Fig.\ \ref{fig:hhgh2}(a). The spectrum is much weaker than the second plateau from H$_{2}$. 

Presumably, the harmonic response from H$_{2}^{+}$ is substantially
enhanced by the action of the oscillating dipole formed by the
recolliding first electron ejected from the neutral molecule and the
neutral ground state. This mechanics is similar to enhancement by an
assisting harmonic pulse \cite{Ishikawa2003, Ishikawa2004,
Takahashi2007, Ishikawa2009}, but the enhancement is due to direct
Coulomb force from the oscillating dipole, rather than harmonics emitted
from it. In the words of the semiclassical three-step model, the
recolliding first electron virtually excites H$_{2}^{+}$, facilitating
second ionization. Thus, electron-electron interaction plays an
important role in high-harmonic generation in some cases {\color{black}(see also \cite{Yuan2015})}, whereas HHG is usually considered as a predominantly single-electron process.


\begin{figure}[tb]
\includegraphics[width=8cm, bb = 0 0 344 534]{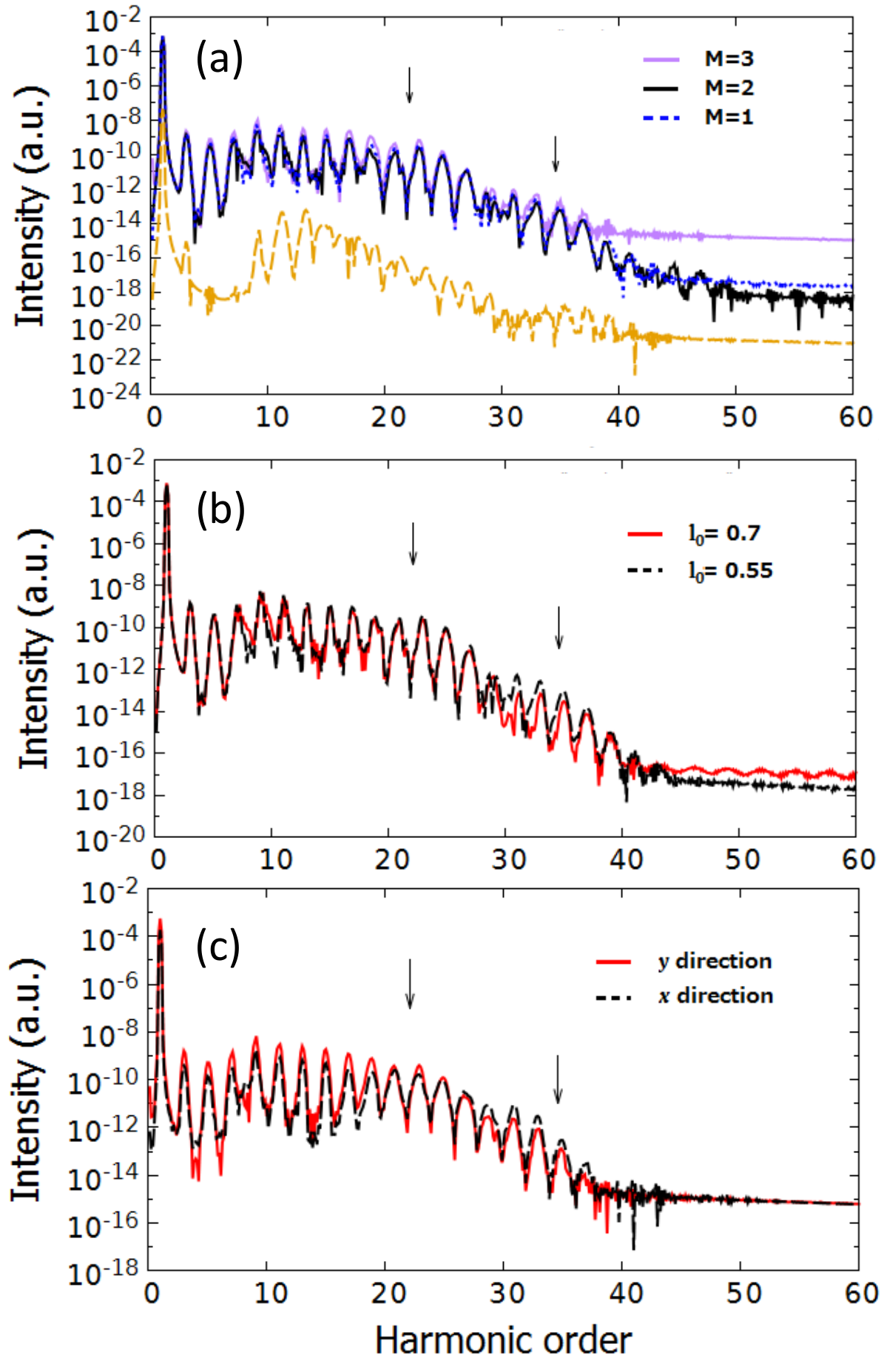}
\caption{
{\color{black}{} Calculated high-harmonic spectra from a hydrogen molecule. See text for laser parameters.
(a) Comparison of the results with $M=1$(blue-dotted), $M=2$ (black-solid), and $M=3$(purple-solid), for laser polarization parallel to the molecular axis. Yellow-dashed line: spectrum from H$_{2}^{+}$ multiplied with the ionization probability of H$_{2}$ ($2.4\times 10^{-4}$).
(b) Comparison of the results with $l_0=0.7$ (red-solid) and 0.55 (black-dashed). The calculation was done with $M=1$ for polarization parallel to the molecular axis. 
(c) Result for laser polarization 30 degrees from the molecular axis [polarization direction is $(\cos 30^\circ, \sin 30^\circ, 0)$]. $M=3$ was used. Red-solid: harmonics emitted in the $y$ direction, black-dashed: in the $x$ direction. Arrows in each panel indicate the cutoff positions expected for H$_{2}$ (22.1-th order) and H$_{2}^{+}$ (34.6-th order). 
}}
\label{fig:hhgh2}
\end{figure}

\subsection{HHG from a water molecule}
\label{sec:water}
As an example of application to molecules of lower symmetry, we simulate the HHG from a water molecule 
{\color{black}{} with its  
oxygen atom located at the origin and two hydrogen atoms at ($\pm
1.4299, 1.10718,0)$.} The side length of cell is set to be 0.6 ($r_{0}>4$), 0.3 ($2<r_{0}<4$) and
0.15($r_{0}<2$) respectively where $r_{0}$ is the distance from the
nearest atom. The outer boundary $x_{L},y_{L},z_{L}$ is set to be 60 (axis
parallel to the polarization) and 30 (axis parpendicular to the
polarization) and the absorption boundary $x_{0},y_{0},z_{0}$ is set to be 0.7
times as long as the outer boundary. The time step size $\Delta t$ is set to be 0.0025.
We use the same laser pulse shape as in Sec.~\ref{sec:heb}. {\color{black} The cutoff energy predicted by the semiclassical three step model is 37.3\, \mbox{eV}, which corresponds to the 12.0 th order}.

Figure\ \ref{fig:hhgwat}, which presents the harmonic spectra for
three different directions of laser polarization, demonstrates high
flexibility of the multi-resolution Cartesian-grid MCTDHF
implementation. {\color{black} One can see that the curves obtained with $M$=5 and 6
almost overlap with each other.
The simulation with $M=6$ took ca. 28 days on a single node with two
hexa-core 3.33 GHz Xeon processors. In this case, the computational
bottleneck was the solution of Poisson's equation (Step 3 of Sec\ \ref{sec:implementation}). 
We expect that the distributed parallelization of the code will
substantially reduce
the computational time, and the extension to TD-CASSCF \cite{Sato2013}
and TD-ORMAS \cite{Sato2015} methods will further extend the
applicability to larger systems.}

\begin{figure*}[tb]
\includegraphics[width=16cm, bb = 0 0 721 394 ]{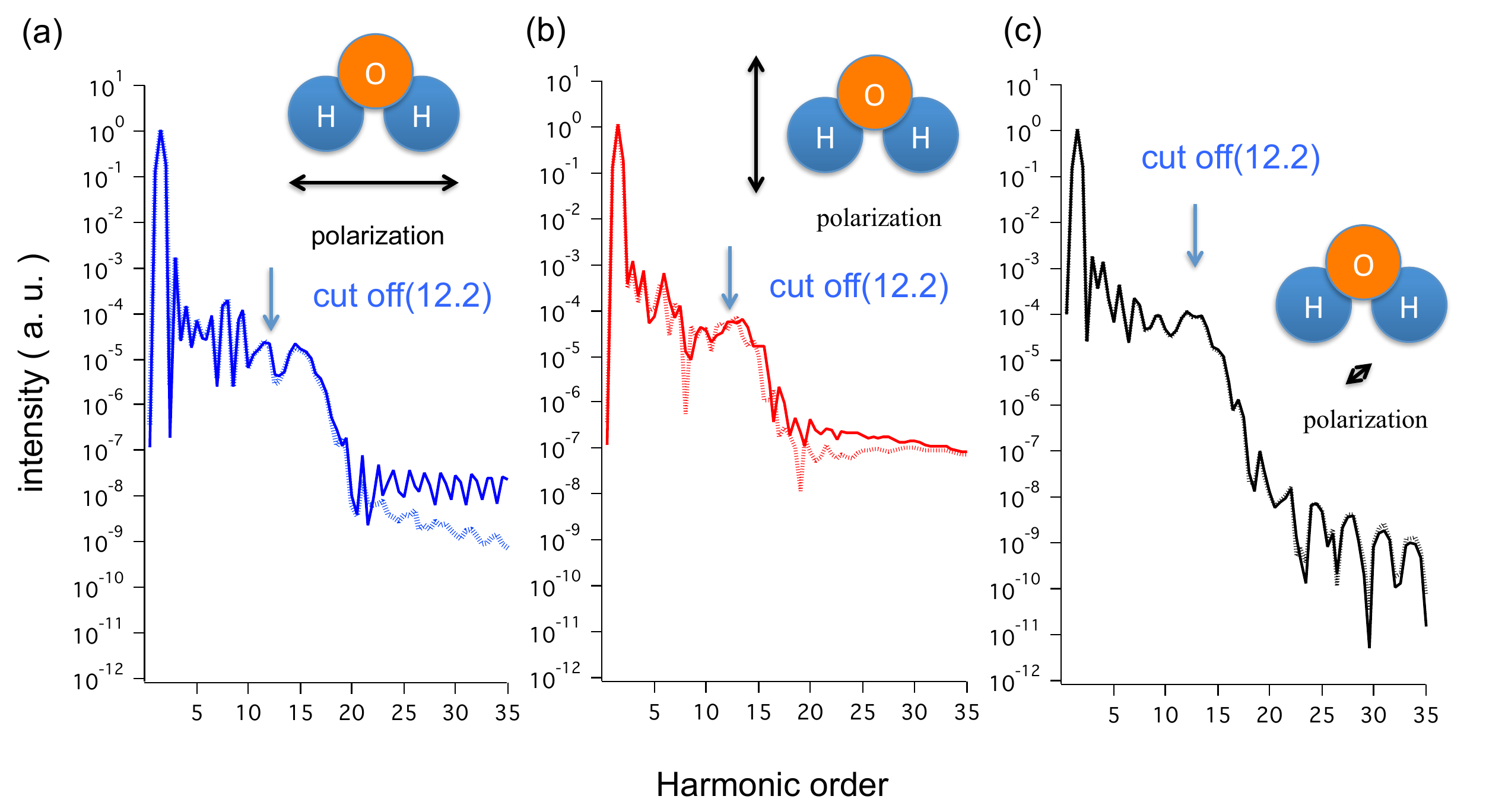}
\caption{{\color{black}{} High-harmonic spectra from a water molecule,
 calculated with {\color{black} $M=5$(dashed) and $M=6$(solid)}, for
 laser polarization along (a) the $x$ axis (b) the $y$ axis (c) the $z$
 axis, as indicated in each panel. Laser polarization in (c) is perpendicular to the plane of the molecule. See text for laser parameters.}}
\label{fig:hhgwat}
\end{figure*}

\section{Conclusion}
\label{sec:conc}

{\color{black}{} We have numerically implemented the MCTDHF method on a
multi-resolution Cartesian grid. Whereas previous approaches have relied
on the underlying symmetries of the simulated atoms and molecules, the present implementation offers a flexible framework to describe strong-field and attosecond processes of real general molecules. Extension to computationally more compact methods such as TD-CASSCF \cite{Sato2013} and TD-ORMAS \cite{Sato2015} will be rather straightforward and enable application to large molecules.

As demonstrations, we have successfully calculated high-harmonic spectra from He, ${\rm H}_2$, and ${\rm H}_2{\rm O}$. As the presence of the second plateau in Fig.\ \ref{fig:hhgh2} implies, the present implementation will uncover yet unexplored multi-electron, multi-channel, and multi-orbital effects, which only first-principles simulations can reveal.}


\begin{acknowledgments}
This work was supported in part by Japan Society for the Promotion of Science (JSPS) KAKENHI Grants No.25286064, No. 26390076, No. 26600111, and No. 26-10100.
This research was also partially supported by the Photon Frontier
 Network Program of the Ministry of Education, Culture, Sports, Science
 and Technology (MEXT) of Japan, the Advanced Integration Science
 Innovation Education and Research Consortium Program of MEXT, the
 Center of Innovation Program from Japan Science and Technology Agency
 (JST), and Core Research for Evolutional Science and Technology, Japan Science and Technology Agency (CREST,
JST). 

\end{acknowledgments}


\bibliography{mctdhf1223}

\end{document}